\documentclass[12pt]{article}
\usepackage{graphicx}
\begin{document}
\centerline{\bf Evolution of ethnocentrism on undirected and directed Barab\'asi-Albert networks}

\bigskip
\centerline{F.W.S. Lima$^1$, Tarik Hadzibeganovic$^{2,3}$ and Dietrich Stauffer$^4$,}

\bigskip
\noindent
$^1$Departamento de F\'{\i}sica,
Universidade Federal do Piau\'{\i}, 64049-550, Teresina-PI, Brazil \\
$^2$Unitat de Recerca en Neuroci\`encia Cognitiva, Departament de Psiquiatria i Medicina Legal, Facultat de Medicina, IMIM-Hospital del Mar, Universitat Aut\` onoma de Barcelona, E-08003 Barcelona, Euroland \\
$^3$Cognitive Science Section, Department of Psychology, University of Graz, A-8010 Graz, Euroland \\
$^4$Institute for Theoretical Physics, Cologne University, D-50923 K\" oln, Euroland. \\
\medskip
{\footnotesize E-mail: fwslima@gmail.com, ta.hadzibeganovic@uni-graz.at, dstauff@thp.uni.koeln.de}
\bigskip

\noindent
{\small {\it Abstract}. Using Monte Carlo simulations, we study the evolution of contigent 
cooperation and ethnocentrism in the one-move game. Interactions and 
reproduction among computational agents are simulated on {\it undirected} and {\it directed} 
Barab\'asi-Albert (BA) networks. We first replicate the Hammond-Axelrod model of in-group 
favoritism on a square lattice and then generalize this model on {\it undirected} 
and {\it directed} BA networks for both asexual and sexual reproduction cases. 
Our simulations demonstrate that irrespective of the mode of reproduction, 
ethnocentric strategy becomes common even though cooperation is individually costly 
and mechanisms such as reciprocity or conformity are absent. 
Moreover, our results indicate that the spread of favoritism toward similar others highly 
depends on the network topology and the associated heterogeneity of the studied population.
}
\medskip

\noindent
{\small {\it Keywords}: Monte Carlo simulation, complex networks, in-group favoritism, ethnocentrism, 
agent-based model, evolutionary model.
\medskip

}
\bigskip

{\bf Introduction}

\noindent
Social phenomena among humans have been modelled extensively, also by physicists
\cite{fortunato}. 
Scale-free networks have recently been claimed as effective promoters of cooperation 
in a variety of social dilemmas \cite{santos}, especially in the prisoner's dilemma game \cite{chess,weibull} 
which has been one of the most widely applied games for studying how cooperative behavior emerges among unrelated 
individuals \cite{axelbook}. An important feature of scale-free networks causing sizeable effects in the evolution 
of cooperation is the heterogeneity of links among individual agents \cite{santos,pacheco}. 

Recent research effort has been directed towards understanding how local interactions between 
computational agents in the prisoner's dilemma game lead to clustered in-group favoritism 
characterized by contingent altruism and cooperation among similar agents \cite{hammond}, as well as 
noncooperation with out-groups and the emergence of 
global ethnocentrism \cite{ar}, even when cooperation is individually costly and the necessary 
mechanisms such as reciprocity \cite{schnegg} and direct self-interested gain \cite{hamilton,brewer}, leadership \cite{zi}, 
reputation \cite{nowak}, trust \cite{hardin}, or conformity \cite{simon} are missing. 

Hammond-Axelrod (HA) models of evolution of ethnocentrism \cite{ar} and contingent altruism \cite{hammond} 
show that in-group favoritism can evolve under a wide variety of conditions, 
even when there is no bias towards apparently similar agents. Instead, four different 
types of agents (each labeled with a different color) populate a 
simple square lattice and compete for limited space via prisoner's dilemma type interactions. "Ethnocentric" 
agents treat other agents within their group more beneficially than those outside of the group and 
in addition, a mechanism for inheritance (genetic or cultural) of strategies is included.

A remarkable outcome of these evolutionary model studies is that after a sufficient number of iterations, 
in-group favoritism emerges as a common strategy even though there is no built-in 
mechanism by which agents can recognize and favor their similar others. Moreover, the 
emerging ability to discriminate between the 
in-group and the out-groups on the basis of different colors and local interactions was actually 
shown to overcome egoism and promote cooperative behavior, even when cost for cooperation is 
relatively high and the same-colored defectors that exploit cooperators need to be supressed. 
Thus, in situations where cooperation among similar individuals is especially costly, ethnocentric behavior 
seems to become necessary to sustain cooperation \cite{ar}. 

However, a significant limitation of these studies \cite{hammond,ar} is that they have 
addressed the evolution of in-group favoritism only within the context of simplified square-lattice-based 
models, neglecting thus the fact that a plethora 
of biological, social, and technological real-world networks of contacts 
are complex and mostly heterogeneous in nature (e.g. scale-free or broad-scale networks) \cite{santos,amaral,dorogotsev,ba2}. 
In addition, the populations studied in \cite{hammond} and \cite{ar} were assumed to reproduce only asexually. 
However, sexual reproduction \cite{staucebrat} in computational models is not just more realistic but it also allows 
for simulation of 'mixed marriages' between agents of different 'ethnicities'. 

It is well-known that in many racist societies, besides using in-group ideology as a control mechanism \cite{buchanan}, 
there were laws against mixed marriages because they disturb ethnocentricity. 
Thus, by incorporating the sexual reproduction in the HA model, it might be possible to investigate how, as a function 
of the probability of mixed marriages, the emergence of ethnocentricity is first made more difficult 
and then eventually, totally prevented.

Another concern of the present work was that, with just a few notable 
exceptions (e.g. \cite{newman,sanchez}), previous computational studies of complex 
networks have largely focused on undirected network systems, even though many 
real-world network structures such as the transcriptional regulatory 
network of the budding yeast (Saccharomyces cerevisiae) or Google's web pages, are actually directed \cite{palla}. 

Our objective here is therefore to generalize the HA model of ethnocentrism 
by including the sexual mode of reproduction among computational agents and 
by performing simulations on both {\it undirected} and {\it directed} scale-free 
networks \cite{ba2,ba1,alex,sumour,sumourss,lima}. More specifically, we first replicate the HA model on a square 
lattice where only asexual reproduction among agents is allowed. Next, we include the sexual mode of reproduction in 
the standard lattice-based HA model. Finally -- and this is the novel emphasis of the present work -- we 
extend the HA model by simulating both reproduction modes separately on {\it undirected} 
and {\it directed} Barab\'asi-Albert (BA) networks. These complex networks have been 
studied extensively by Lima et al. in the context of magnetism \cite{lima0,lima1,lima2,lima3,lima4,lima5,lima6} 
and econophysics models \cite{lima7,lima8}. 

While acknowledging the possibility of individual differences in cooperative behavior \cite{africanlions}, 
we are interested here in how ethnocentric behavior, defined as preferential in-group favoritism 
and lack of cooperation with the out-groups, emerges spontaneously at the global group-level 
from the local interactions of differently labeled 
individual agents. Since our model is not analytically tractable, we employ extensive 
agent-based Monte-Carlo simulations, detailed in the next section.

\bigskip

{\bf Model and Simulations}

\noindent
Here, we study the effects of network structure and mode of reproduction (both asexual and sexual, respectively) 
in the standard (lattice-based) and extended (scale-free network-based) HA model of ethnocentric behavior. 
Ethnocentrism is operationalized as preferential in-group cooperation and non-cooperation with 
any out-group members. Furthermore, cooperation is set to be an individually costly endeavour, 
due to implementation of a one-move game
dilemma framework. Moreover, any direct reciprocity is disabled by employing a one-move 
game rather than the iterated version. 

In order to establish group differences and to enable their potential detectability, each agent 
is labeled by three traits: a) color, b) strategic behavior when interacting with same-colored agents, 
and c) strategic behavior when interacting with differently colored agents. 

A color specifies the group membership of an agent and is interpreted as an observable feature that may be seen 
as socialy relevant in a given population (e.g., skin color, religion, political orientation, or language). 
An "ethnocentric" agent is defined as one that cooperates with same-colored agents, but does 
not cooperate with individuals of different color. 

Thus, the "ethnocentric" in-group preference and discrimination against the out-groups is only one out 
of four possible strategies. An "altruist" cooperates with all agents, while an "egoist" always defects. 
A "cosmopolitan" agent cooperates with agents of a different color but not with those of the same color. 
Given the fact that colors and strategies are not linked, the model allows for the existence of defectors that always 
exploit the cooperators. Thus, they are receiving help from same-colored ethnocentric agents 
and at the same time, provide help to no one at all \cite{ar}. 

In the present paper, each simulation begins with an empty lattice or an empty network. Generally, at each time step, 
the following events occur:

\bigskip
a) Asexual reproduction mode: 

1. New agents with random traits invade at random empty lattice/network sites.

2. All agents start with an initial chance of reproducing, i.e., their initial 
potential to reproduce (iPTR). Each pair of adjacent agents interacts in a 
one-move game in which each agent selects whether or not to provide help to 
the other. As a result, agents can either gain, or lose some of their iPTR.

3. In random order, each agent is selected and given a chance to reproduce. An offspring is then 
created inheriting the traits from its parent, with a specific mutation rate per each trait. 
Agents are cloned only if there is an empty space next to them. Each agent's birth-rate 
is reset to the iPTR.

4. Population dynamics is further determined by a fixed mortality rate, making thus 
room for future offspring and new immigrants invading at subsequent iterations.

5. The steps 1-4 are iterated until the total simulation time is finally reached.

\bigskip
b) Sexual reproduction mode:

For the sexual case, the process is generally the same as in the asexual reproduction mode, 
except for the step 3 where in the reproduction process we include two new traits: One for sex (male/female) 
and the other one for the scope of reproduction, allowing sex only among 
non-relatives (thus, reproduction between brothers and sisters is forbidden).

\bigskip
In all simulations reported in this study (with both asexual and sexual cases), we used networks with a total 
of 1002001 sites and evolved the system over $3 \times 10^{6}$ time steps. 
When the final iteration step was reached, the four types of agents (ethnocentric, altruist, egoist and cosmopolitan) 
were counted and compared. All simulations were written in FORTRAN and conducted on an 
UNIX-based SGI Altix 1350 Cluster server. Simulations always started with an empty space 
of 1002001 sites for all networks. The space was helical (for square lattices), with every site having exactly four 
neighboring sites. One time step was one update attempt per lattice/network site consisting of four 
different stages: immigration, interaction, reproduction, and death. 

\bigskip
More specifically:

1. At stage 1, an immigrant with random traits enters the network and occupies at a random an empty site.

2. The iPTR of each agent is set to 12 percent. Each pair of neighbors
then interacts in a one-move game. Giving help has a cost, namely, a decrease in the agent's 
PTR by 1 percent. On the other hand, receiving help has a benefit, namely, an increase in the agent's 
PTR by 3 percent.

3. Each agent is selected in a random order and given a chance to reproduce with 
probability equal to its current PTR. Reproduction consists of creating and placing an offspring in 
an adjacent empty site, if there is one. If there is no empty space, no new agents are created. 
Furthermore, an offspring inherits the traits of its parent, with a mutation rate of 0.5 percent per trait.

4. Finally, each agent has a 10 percent chance of dying, making thus room for future offspring and immigrants. \\

\bigskip

{\bf Results}

Our results are summarized in Table 1 and Figs.1-3. The main finding is that the ethnocentric strategy becomes clearly dominant
on square lattices irrespective of the involved mode of reproduction. Slightly over $77\%$ 
of all agents adopt the ethnocentric strategy after $3 \times 10^{6}$ time steps. 

Surprisingly enough, for the {\it directed} BA network case, the results are again roughly 
the same for both modes of reproduction, but now all different strategies are 
roughly equally distributed, with $\simeq 25$ percent per strategy. 

Finally, the  {\it undirected} BA network simulation with the asexual mode of reproduction yielded again the 
dominance of the ethnocentric strategy. There were $\simeq 71$ percent of ethnocentrics and $\simeq 20$ 
percent of egoists. 

However, when sexual reproduction among agents on an undirected BA network was allowed, 
we found a stronger competition between the ethnocentrics and egoists: The simulation 
yielded $\simeq 43$ percent of ethnocentrics and $\simeq 41$ percent of egoists. Here, we 
demonstrated that the spread of ethnocentrism indeed may become more difficult when agents 
can reproduce sexually, however, this depends on the topology of the involved network.

\begin{table}[htb]
\begin{center}
\begin{tabular}{|c c c c c c c|}
\hline
\hline
$  asexual $ & $SL$     & $BAD$     & $BAUND$ \\
\hline

$ ethnocentrics $  & $ 77.31$ & $25.60 $  & $ 70.53$ \\ 
$ altruists   $     & $ 8.19 $ & $25.53 $  & $ 0.45 $ \\
$ egoists $        & $ 9.32 $ & $24.37 $  & $ 20.35 $ \\
$cosmopolitan $    & $ 5.15 $ & $24.48 $  & $ 0.45 $ \\

\hline
$  sexual $   & $SL$     & $BAD$     & $BAUND$ \\
\hline

$ ecthnocentrics $ & $  77.28 $ & $25.50  $  & $ 42.56$ \\ 
$ altruists   $    & $   8.43 $ & $25.55  $  & $  4.50 $ \\
$ egoists $       & $   9.00 $ & $24.57  $  & $ 41.11 $ \\
$cosmopolitan $   & $   5.27 $ & $24.36  $  & $ 11.36 $ \\

\hline
\hline
\end{tabular}
\end{center}
\caption{Evolution of ethnocentrism on square lattices (SL), 
{\it directed} (BAD) and {\it undirected} (BAUND) Barab\'asi-Albert 
networks for both modes of reproduction. Distribution of four strategies shown in percent.} \label{table1}
\end{table}

{\bf Conclusion}

We have investigated the evolution of ethnocentric behavior on scale-free networks which are, 
as the majority of most natural networks of contacts, heterogeneous in nature \cite{santos,amaral,dorogotsev,ba2}. 
Moreover, we have replicated the standard HA model of ethnocentrism and then studied how in-group 
favoritism emerges on {\it undirected} and {\it directed} BA networks, when both asexual 
and sexual reproduction modes are enabled. For the undirected BA network simulation, we showed
that the percolation of ethnocentrism becomes more difficult when agents can reproduce sexually, 
and in such a case, ethnocentrics and egoists compete for space more intensively. However, we 
also demonstrated that this result strongly depends on the topology of the simulated network.

Future studies should take into account more fine-granularity with respect to the interaction 
and reproduction processes among computational agents \cite{goldstone}. For example, one could 
investigate how divorces in mixed marriages (e.g. ethnocentric-altruist) may contribute to the 
survival of ethnocentric behavior once ethnocentricity becomes "endangered" in a given population, 
or, how divorces of pure "ethnocentric couples" may contribute to the eventual extinction of 
pronounced in-group favoritism. Furthermore, one should consider that there is always a certain 
fraction of individuals who never change their strategy throughout their 
lifetime (or throughout the simulation). 

As suggested in \cite{buchanan}, future model generalizations should also consider simulating 
more realistic cases of ethnocentrism evolution. For instance, extended models of 
ethnocentrism \`a la Hammond \& Axelrod might 
further help in understading issues previously raised in the context of language 
competition and opinion dynamics motivated by inter-ethnic differences \cite{bosnia,nyas}. 
Moreover, further biological model \cite{hammond} 
extensions on BA, Erd\" os-R\' enyi \cite{erdos}, and small-world \cite{ba2} networks should be developed. 

In sum, we showed that the spread of ethnocentrism can become an important mechanism for promoting high 
levels of costly cooperation but can be reduced when different types of agents are allowed to mix and 
sexually reproduce. However, this effect of sexual reproduction is confounded with the effect of network topology. \\

The authors thank the Brazilian agency FAPEPI (Teresina-Piau\'{\i}-Brasil) for financial support. This
work was also supported by SGI\textregistered Altix\textregistered 1350 and the computational park
CENAPAD-UNICAMP-USP, SP-BRAZIL.

\begin{figure}[hbt]
\begin{center}
\includegraphics[angle=0,scale=0.41]{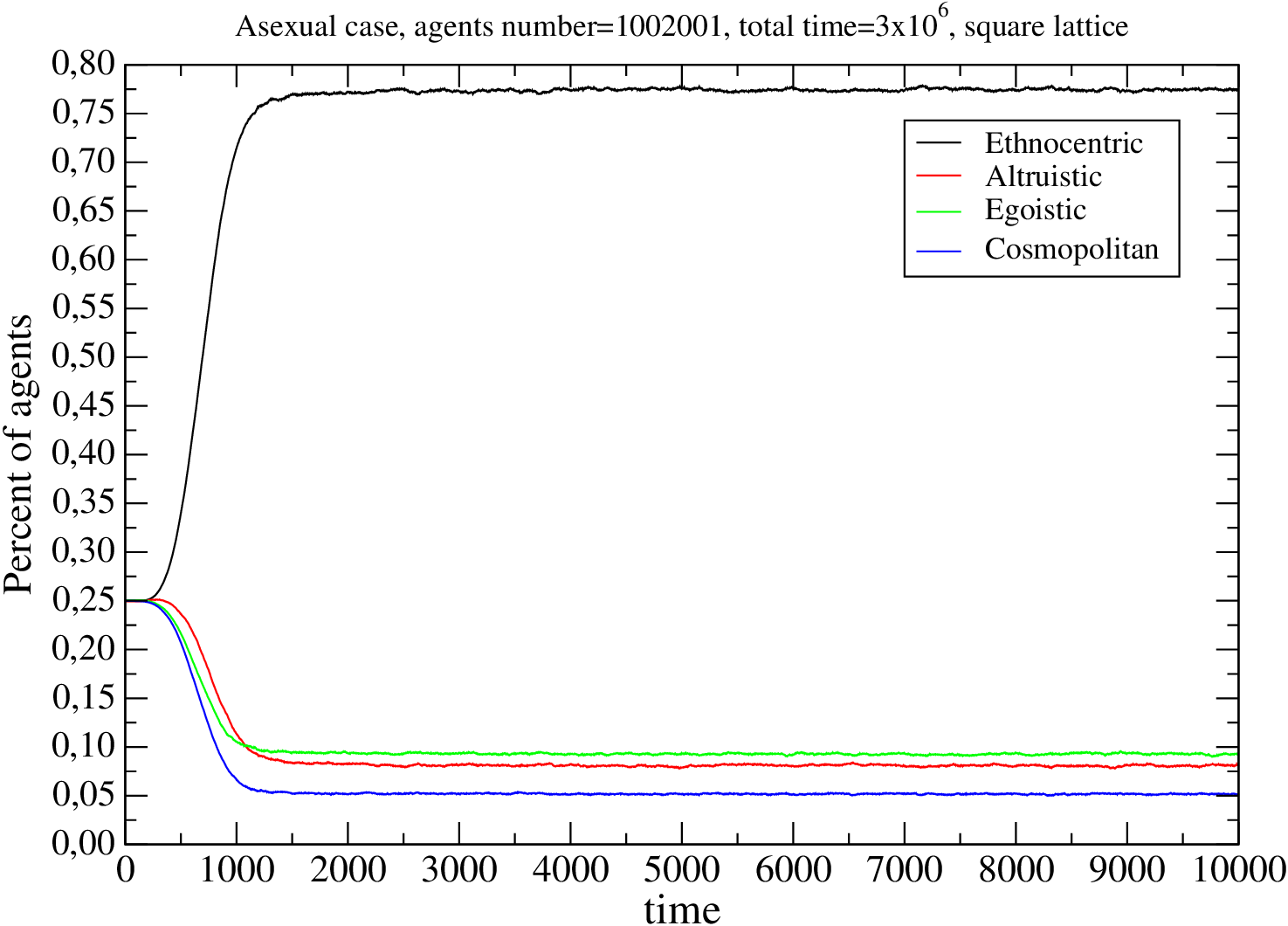}
\includegraphics[angle=0,scale=0.41]{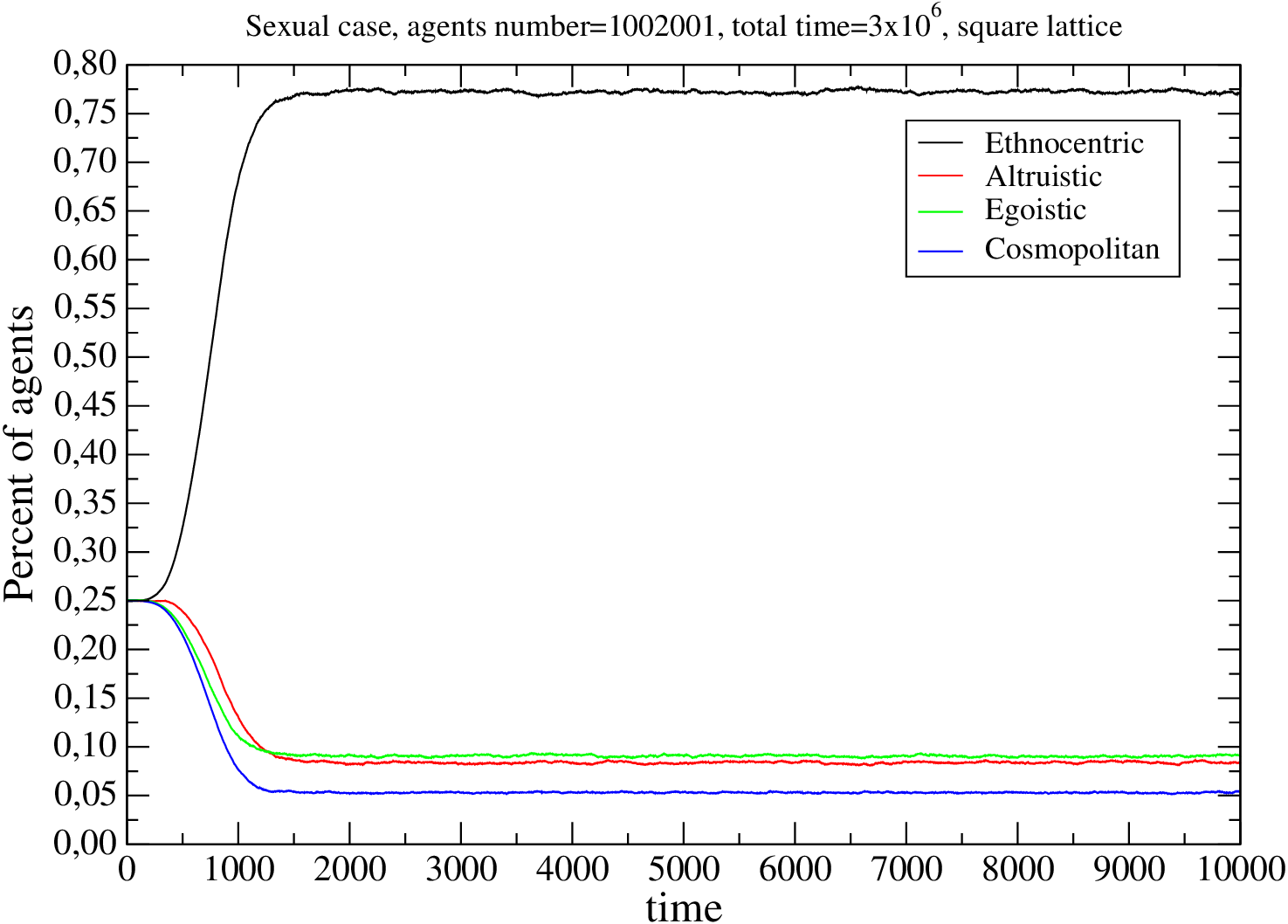}
\end{center}
\caption{Percentage of agents versus time for ethnocentric (black), altruistic (red), egoistic(green) 
and comospolitan (blue) agents on square lattices of size $L=1001$ with $N=L \times L=1002001$ agents 
and time of $3 \times 10^{6}$ iterations for asexual case (top) and sexual (bottom). Here, and in Fig.3, the "altruistic curve" is 
above the "cosmopolitan curve".}
\end{figure} 
\bigskip

\begin{figure}[hbt]
\begin{center}
\includegraphics[angle=0,scale=0.41]{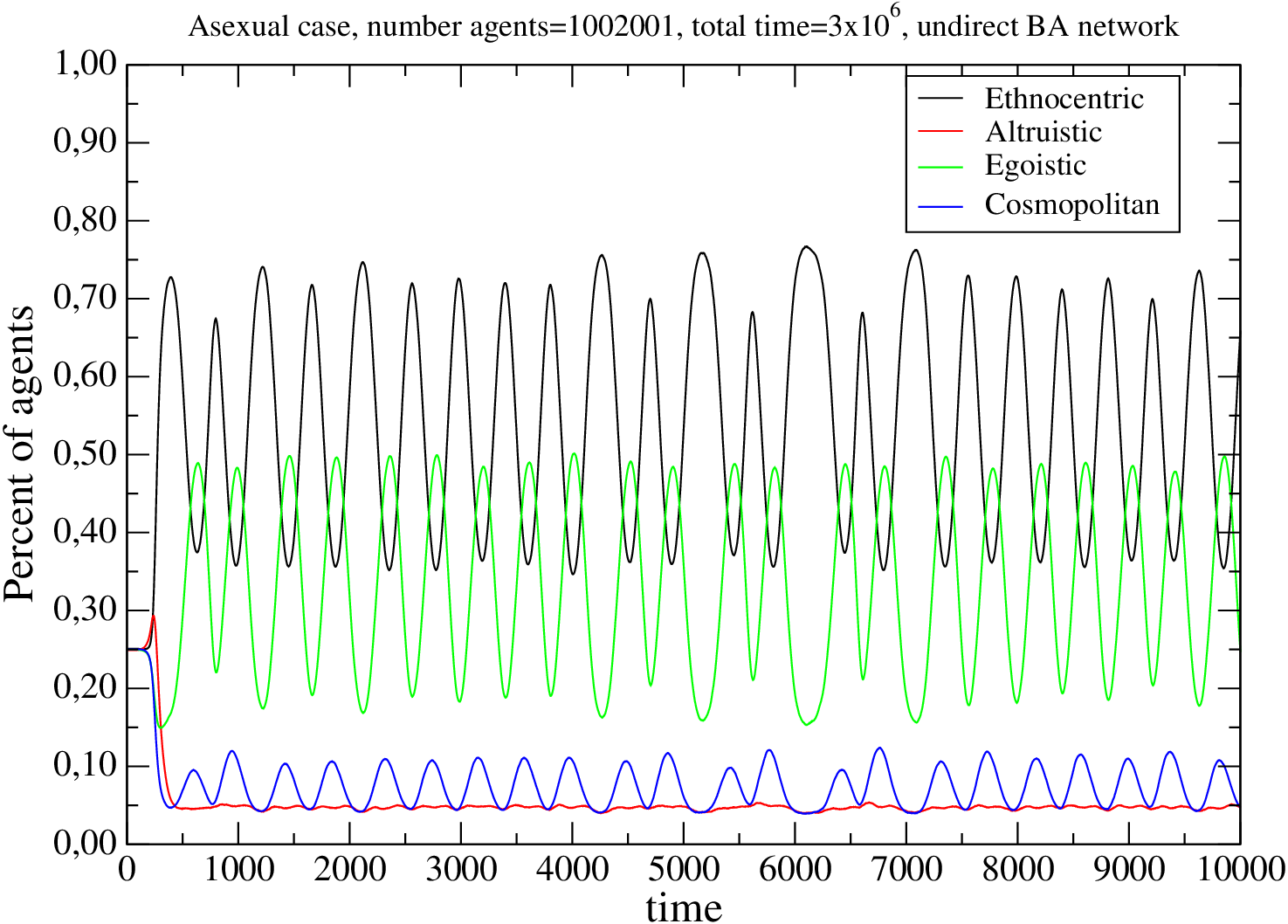}
\includegraphics[angle=0,scale=0.41]{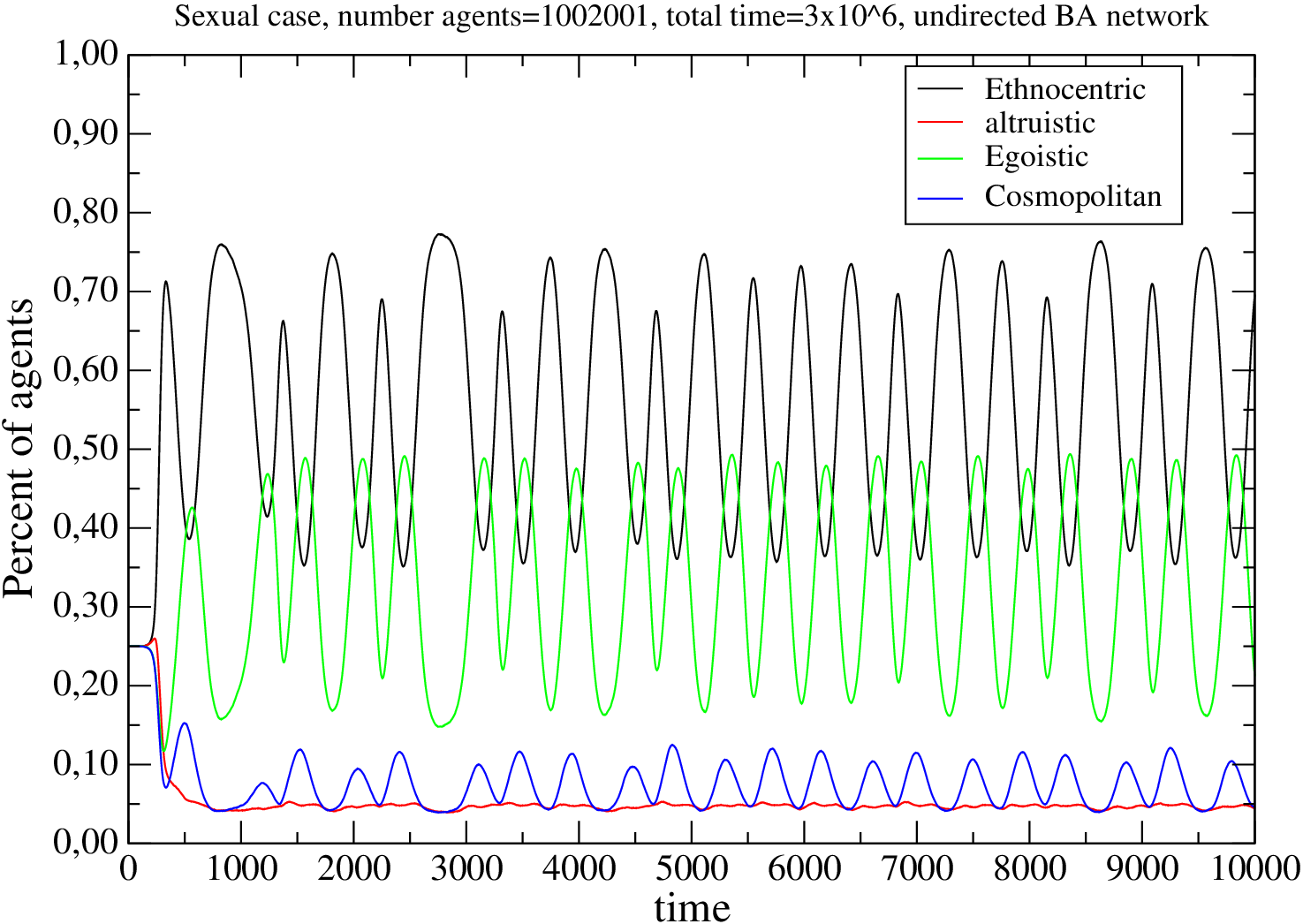}
\end{center}
\caption{As before, results for $N=1002001$, but now for {\it undirected} Barab\'asi-Albert(BA) asexual case 
(top) and sexual case (bottom). Altruists show nearly no oscillations.}
\end{figure}
 
\begin{figure}[hbt]
\begin{center}
\includegraphics[angle=0,scale=0.41]{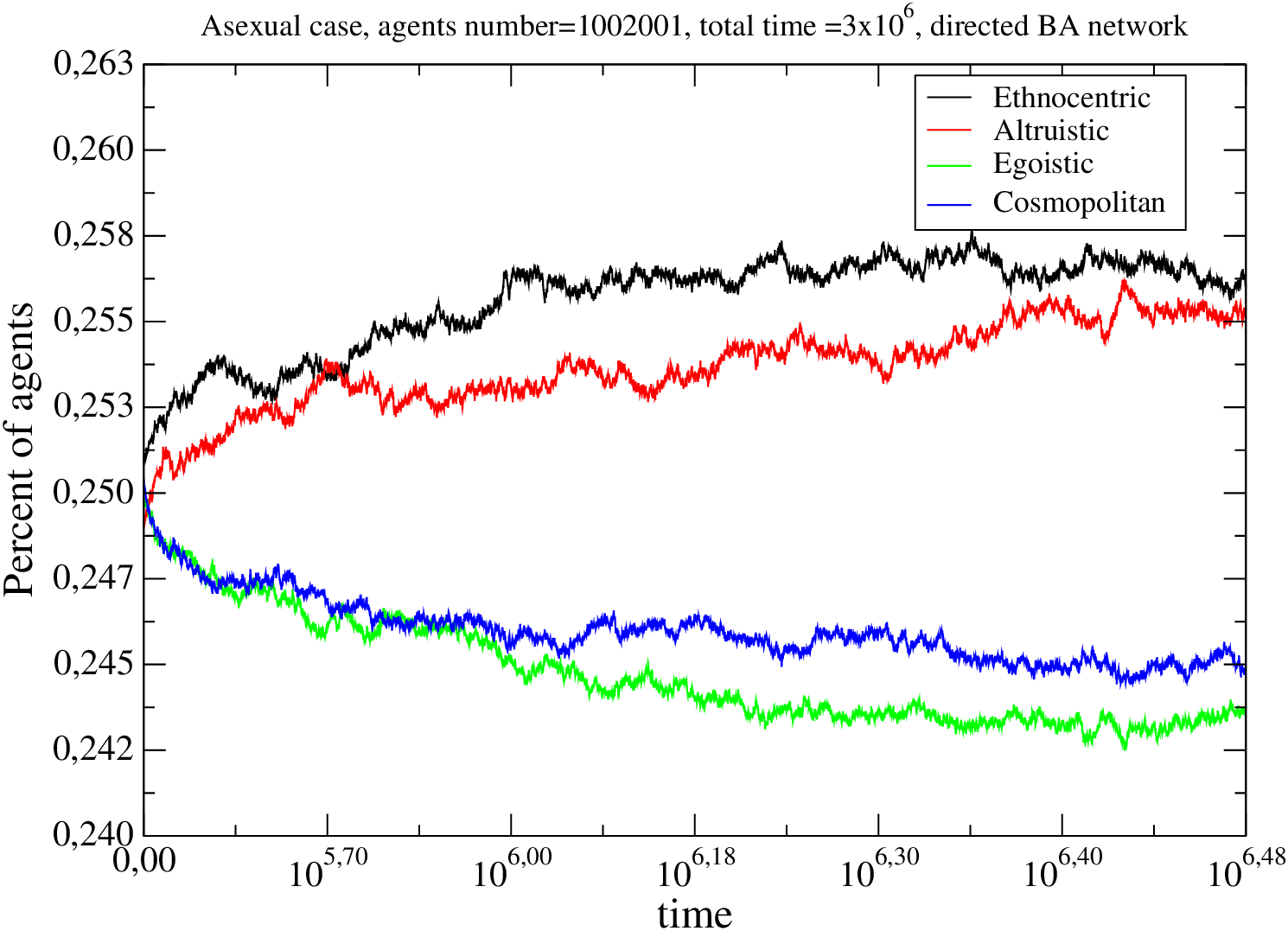}
\includegraphics[angle=0,scale=0.41]{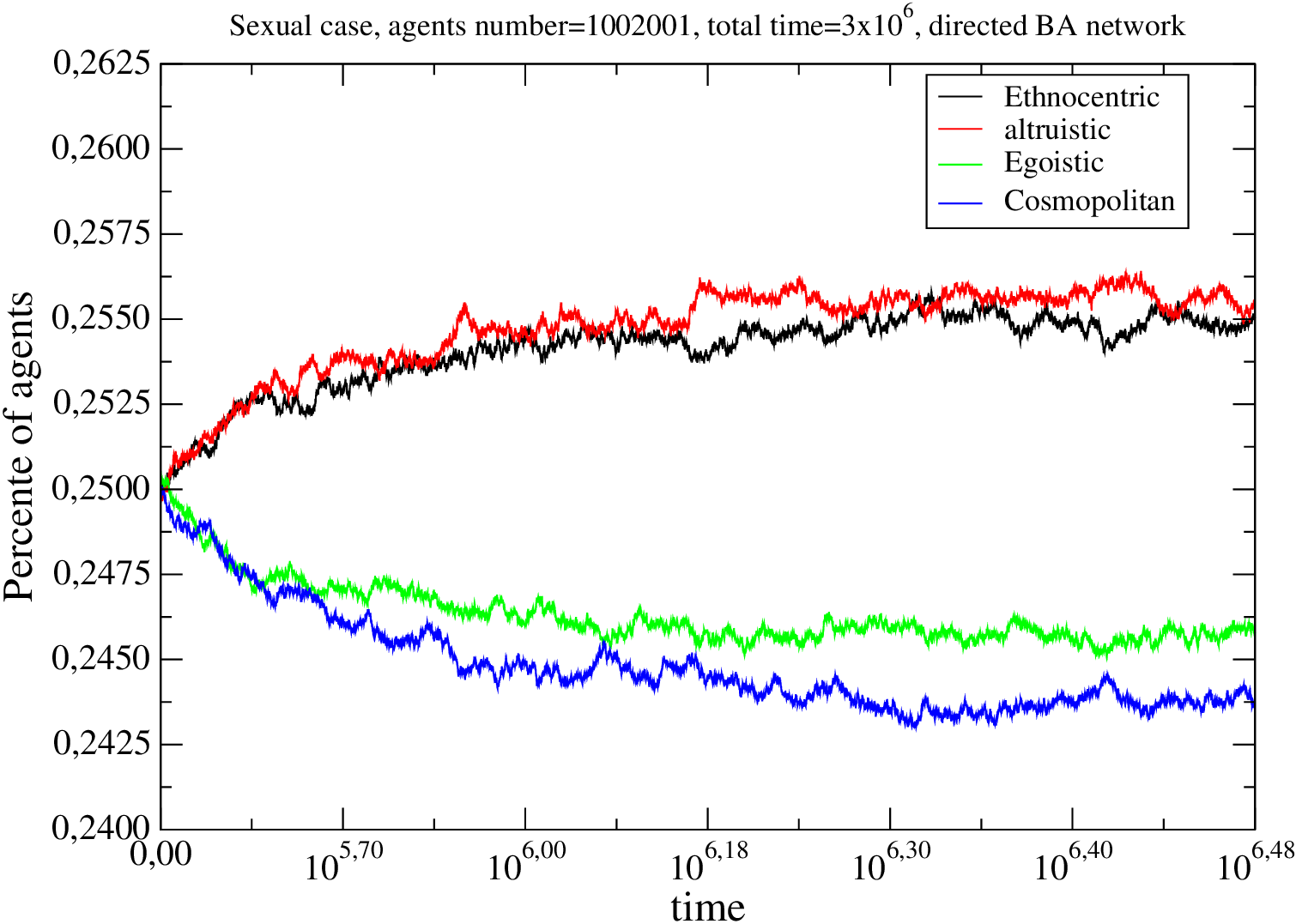}
\end{center}
\caption{As before, but now for {\it directed} BA networks, again asexual case (top) and sexual case (bottom). } 
\end{figure}

\bigskip

\end{document}